\documentclass[reprint,
superscriptaddress,
 amsmath,amssymb,
 aps,
]{revtex4-2}

\usepackage{graphicx}
\usepackage{dcolumn}
\usepackage{bm}
\usepackage{underscore}
\usepackage[english]{babel}

\usepackage{physics}
\usepackage{placeins}
\usepackage{algorithm2e}
\usepackage{algpseudocode}
\usepackage{cleveref}
\usepackage{siunitx}

\preprint{APS/123-QED}

\begin{document}

\title{ManQala: Game-Inspired Strategies for Quantum State Engineering}

\author{Onur Danaci}
\email{danaci@lorentz.leidenuniv.nl}
\affiliation{IBM-HBCU Quantum Center, Howard University, Washington, DC 20059, USA}
\affiliation{Instituut-Lorentz, Universiteit Leiden, 2300 RA Leiden, The Netherlands}

\author{Wenlei Zhang}
\affiliation{Department of Physics \& Engineering Physics, Tulane University, New Orleans, LA 70118, USA}

\author{Robert Coleman}
\affiliation{IBM-HBCU Quantum Center, Howard University, Washington, DC 20059, USA}

\author{William Djakam}
\affiliation{Department of Electrical \& Computer Engineering, University of Alabama-Birmingham, Birmingham, AL 35233, USA}

\author{Michaela Amoo}
\affiliation{IBM-HBCU Quantum Center, Howard University, Washington, DC 20059, USA}

\author{Ryan T. Glasser}
\affiliation{Department of Physics \& Engineering Physics, Tulane University, New Orleans, LA 70118, USA}

\author{Brian T. Kirby}
\affiliation{Department of Physics \& Engineering Physics, Tulane University, New Orleans, LA 70118, USA}
\affiliation{DEVCOM Army Research Laboratory, Adelphi, MD 20783, USA}

\author{Moussa N'Gom}
\affiliation{Department of Physics, Applied Physics, and Astronomy, Rensselaer Polytechnic Institute, Troy, NY 12180, USA}

\author{Thomas A. Searles}
\email{tsearles@uic.edu}
\affiliation{Department of Electrical \& Computer Engineering, University of Illinois Chicago, Chicago, IL 60607, USA}

\date{\today}

\begin{abstract}
The ability to prepare systems in specific target states through quantum engineering is essential for realizing the new technologies promised by a second quantum revolution. 
Here, we recast the fundamental problem of state preparation in high-dimensional Hilbert spaces as ManQala, a quantum game inspired by the West African sowing game mancala. 
Motivated by optimal gameplay in solitaire mancala, where nested nearest-neighbor permutations and actions evolve the state of the game board to its target configuration, ManQala acts as a pre-processing approach for deterministically arranging particles in a quantum control problem.
Once pre-processing with ManQala is complete, existing quantum control methods are applied, but now with a reduced search space. 
We find that ManQala-type strategies match, or outperform, competing approaches in terms of final state variance even in small-scale quantum state engineering problems where we expect the slightest advantage, since the relative reduction in search space is the least.
These results suggest that ManQala provides a rich platform for designing control protocols relevant to near-term intermediate-scale quantum technologies.
\end{abstract}

\maketitle

\section{Introduction}
\label{sec:intro}

Quantum engineering applies traditional principles of engineering such as design and control to quantum phenomena, devices and systems. In particular, quantum state engineering (the application of control methods to quantum state preparation problems~\cite{d2021introduction, wiseman2009quantum}), or QSE,  is of interest for quantum computing \cite{ramakrishna1996relation, ball2021software}, networking \cite{duan2001long, muller2013room}, and sensing \cite{rembold2020introduction, poggiali2018optimal} applications.  
In general, one can separate QSE into three classes: preparation \cite{domokos1994role,milburn1986dissipative, danageozian2022noisy}, stabilization \cite{rossi2018measurement,   viola1999dynamical} and purification \cite{ticozzi2014quantum, nielsen2010quantum} of quantum states.
Primarily, QSE strategies to prepare a target state in a high-dimensional Hilbert space make use of two techniques, and combinations thereof. 
The first is based on unitary time evolution with respect to some known control Hamiltonian(s). This evolution is deterministic and known as coherent control, or unitary control \cite{wiseman2009quantum,sorensenthesis}. 
The second technique uses measurement back-action to steer the quantum state stochastically and is known as incoherent control, or control-free \cite{hacohen2018incoherent,sorensenthesis}.

Two examples employing both methods in their QSE strategies are FUMES (fixed unitary evolution and measurements)~\cite{pedersen2014many} and Zeno-locked FUMES (Z-FUMES) ~\cite{sorensen2018quantum}.  The FUMES strategy is based on unitarily evolving a state with respect to a known Hamiltonian up to a point where the fidelity between the state in hand and the target state is maximized, $\mathcal{F(\psi,\psi_{\text{targ}}})=\bra{\psi}\ket{\psi_{\text{targ}}}$, then making a probabilistic projective measurement \cite{pedersen2014many}.
The Z-FUMES strategy implements the same search algorithm, but makes use of the quantum Zeno effect \cite{misra1977zeno, ghirardi1979small, itano1990quantum,Nodurft:22}, to ``lock'', i.e. halt the evolution of, certain subspaces of the system to gradually shrink the total search space \cite{facchi2002quantum, facchi2004unification, facchi2008quantum}. 
Example applications of measurement back-action methods include the control of qubits for quantum computing \cite{franson2004quantum, paz2012zeno, dominy2013analysis, hacohen2018incoherent}, in control of quantum optical systems \cite{lee2002linear, Nodurft:22}, and in control of critical behaviour of quantum gases \cite{elliott2016quantum, fuji2020measurement, ivanov2020feedback}.

Recently, parallels between control problems and games have emerged as demonstrated by the application of AlphaZero~\cite{silver2016mastering}, which was initially developed for playing games like Go and chess, to the optimal control of inverted pendulums~\cite{moerland2018a0c}.
Furthermore, both control problems and games are concerned with selecting actions while interacting with some environment to change the system's state; typically based on some rules and signals from the environment \cite{bertsekas2019reinforcement}.  A substantial amount of interest has been focused on using games in quantum information science not only as a pedagogical tool to develop intuition, but also as a legitimate way to solve problems. For example, a quantum version of sudoku, referred to as SudoQ, was found to be closely related to the topic of mutually unbiased bases \cite{paczos2021genuinely}. Many quantum analogues to  games have been studied, such as Go \cite{ranchin2016quantum}, tic-tac-toe \cite{goff2006quantum,leaw2010strategic}, chess \cite{akl2010importance}, blackjack \cite{mura2020quantization}, roulette \cite{wang2000quantum}, and sudoku \cite{nechita2020sudoq,paczos2021genuinely}, with various motivations. In general, a game merges with principles of quantum mechanics by associating game pieces with quantum states and defining operators that evolve the game in analogous ways to the original gameplay. 

Noticeably, a missing link exists between the emerging fields of QSE and quantum games. For example, experimental implementations of quantum games and associated gameplay require extensive control of the underlying quantum systems and rely on QSE. However, as we will focus on in this paper, the reverse relationship is also meaningful: the development of QSE strategies  inspired by games. 

Here, we introduce a quantum version of the traditional West African sowing game mancala, which we refer to as ManQala, and explore its applicability as a framework for state engineering.
We present QSE strategies inspired by the game that match or surpass the performance of traditional unitary or measurement-based strategies. In particular, the advantages persist even when considering small systems and two-site interactions where one could expect the slightest advantage. Importantly, ManQala-inspired strategies show promise at scaling to higher dimensions better than (Z-)FUMES due to small search spaces and the parallelizability associated with their divide-and-conquer approach.
Hence, the strategies presented in this paper represent a potential path toward developing game-based QSE techniques with improved scaling properties compared to leading alternatives.

This paper is organized as follows.  In Section \ref{sec:engineering}, we provide an overview of the ManQala game and relate it to the basic goals of QSE. From these results, we present useful methods for developing game-inspired strategies for QSE and compare and contrast these to the strategies outlined by FUMES and Z-FUMES.
In Section \ref{sec:two_site}, we provide numerical simulations of ManQala, FUMES, and Z-FUMES algorithms for the scenario where the former has the least advantage against the latter. However, clear advantages for ManQala are demonstrated with respect to parallelization and variance.
Finally, Section \ref{sec:conclusion} concludes the paper.

\section{ManQala Quantum State Engineering Strategy}
\label{sec:engineering}
\subsection{Overview}

Mancala is the generic name for a collection of related games played worldwide for over a thousand years. Mancala games consist of pieces called stones (or seeds) and a game board that includes a set of valid locations, referred to as pits, for game piece placement. Additionally, mancala game boards have a special pit, or pits, called the Ruma.  Various rule sets exist, but mancala is generally a turn-based game where players alternate moving, or ``sowing," stones in counter-clockwise or leftward direction from pit-to-pit in a series of chained operations to collect the stones in a Ruma. 

This Section develops techniques to play quantum mancala, or ManQala, that can be applied to any variation of the traditional game, including multiplayer iterations. However, we have found that even the most basic solitaire versions of mancala have a rich enough game structure to reveal significant and uniquely quantum features of ManQala. For this reason, throughout the rest of this paper, we will focus on Tchoukaillon, a solitaire version of mancala which is particularly amenable to mathematical analysis. 

\begin{figure}
    \centering
    \includegraphics[width=\linewidth]{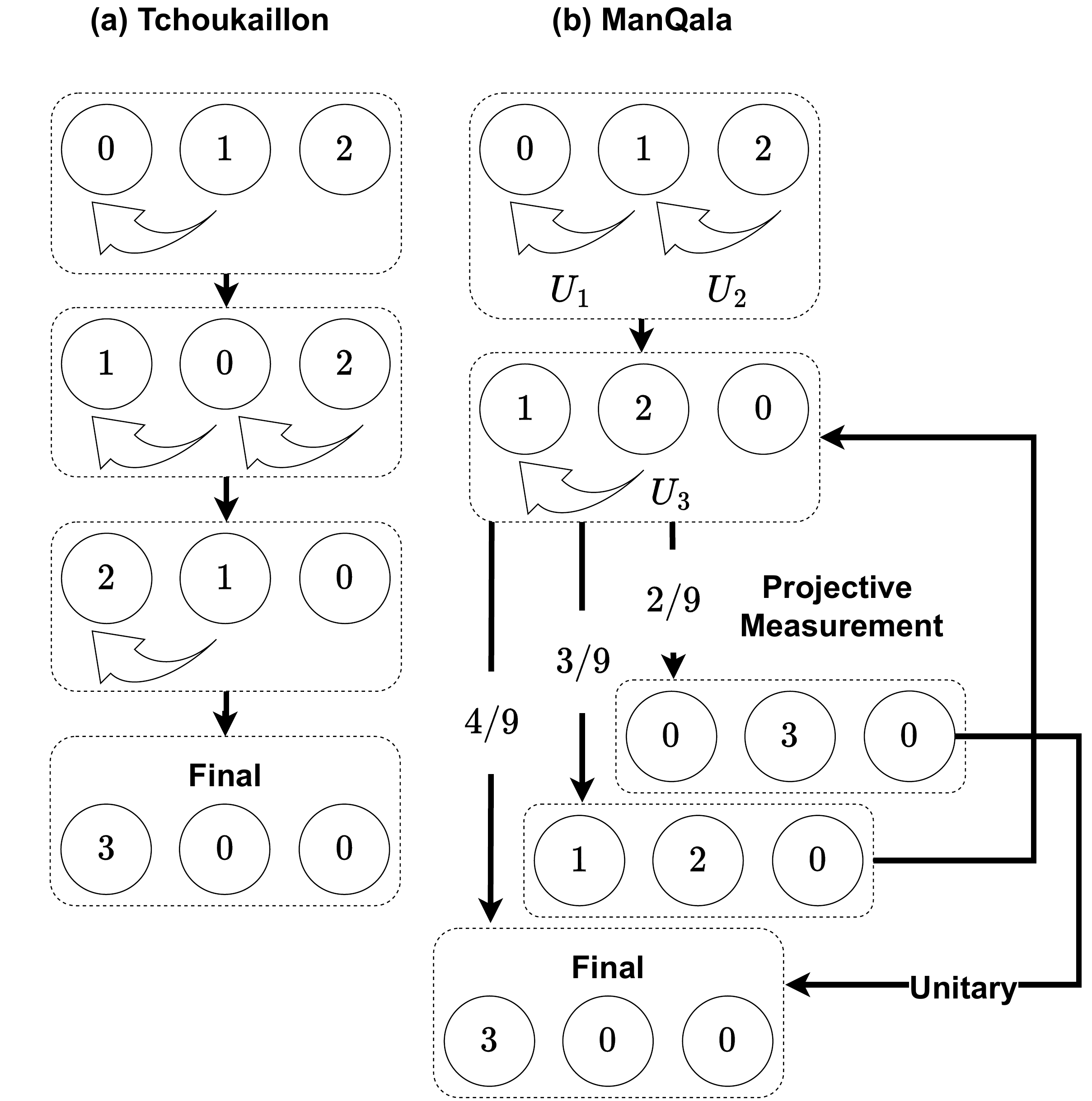}
\caption{\textbf{Example game boards for (a) Tchoukaillon (a solitaire variant of mancala) and its direct quantum analogue ManQala in (b).} Here, we show both boards with $N=3$ stones and $M=3$ lattice sites, and we represent sowing with arrows (which become two-site unitary operators in ManQala). The sequential unitary actions $U_1$ and $U_2$ in the Figure represent the deterministic quantum analogue of the first two Tchoukaillon moves via site-population permutations. The final step of the Tchoukaillon game has no deterministic unitary realization in the quantum version of the game. Hence, $U_3$ drives the state where the probability of observing the winning board is maximized. Upon observation (projective measurement), the target state is achieved with a probability of $6/9$ (where an additional unitary may be required), or the board reverts to the configuration before $U_3$, and the final step is repeated until successful.}
    \label{fig:board}
\end{figure}

Tchoukaillon was developed by Deledicq et al. \cite{deledicq1977wari} in 1977 and is itself a derivative of another, much older mancala variant called Tchouka \cite{sainte1929geometrie}. Game boards in Tchoukaillon are linear and consist of only a single Ruma, which we assume is the left-most pit by convention.   The sowing rules state that when a  player picks $N$ stones up from a pit, they leave that pit empty to sow all those stones to the pit on the left-hand side. Then, they proceed by picking up $N-1$ stones from that pit and sowing them to the next pit on the left, and so on until there are no stones remaining in their hands. In Tchoukaillon,  a player must end each round of sowing by placing the final stone in the Ruma, with under- and over-shooting resulting in losing the game. The win conditions for Tchoukaillon are, in fact, so restrictive that the initial game board ultimately determines whether a win is possible and, if so, prescribes the unique series of moves needed to win \cite{broline1995combinatorics, jones2013solitaire}. 

Assuming we begin with a Tchoukaillon board for which winning is possible, then the series of moves that will result in winning can be described succinctly. In particular, the stones from the pits nearest to the Ruma (empty pit at the edge) need to be picked and sowed first, with this strategy repeated, moving to the next non-empty pit until the game board is cleared.
In Figure \ref{fig:board}(a), we visually present an example of this strategy for a winning board with a Ruma and two additional pits.

The game play in Figure \ref{fig:board}(a) consists of two movements; the first is equivalent to a permutation, such as in the first step when sowing into an empty pit, and the second is the merging movements that combine seeds from two pits. This distinction is significant as we consider a quantum version of the game since permutation can be performed deterministically with unitary evolution, while merging operations are nondeterministic. Notably, permutation movements alone bring the board ``closer" to the final configuration. This observation that deterministic permutation operations can significantly reduce the number of tunneling events needed to achieve the target state and effectively reduce board size will be the primary strategy we will use throughout this paper as we develop ManQala. Furthermore, even as we move away from problems that have direct analogies with mancala, we will see that pre-processing in this fashion reduces the problem space making the application of search-based quantum control approaches (e.g., Z-FUMES) more efficient. 

In this paper, we aim to devise a family of QSE strategies, termed ManQala, that mimic the mechanics of a traditional Tchoukaillon game. 
For context, in Figure~\ref{fig:board}(b) we include a simple example of the same game shown in Figure~\ref{fig:board}(a) but where ``stones'' are replaced with bosonic states and ``pits'' by system modes. This now constitutes a quantum system where we can apply ManQala. In this simple ManQala example, ``sowing'' is performed with unitary operations $U_i$, for $i = 1, 2, 3$.
Even in this basic example, with only three modes and three bosons, the gameplay diverges from mancala in two important ways. 
First, since the quantum version manipulates amplitudes of quantum systems and avoids ``collapse,'' the game does not result in a ``winning'' configuration with certainty.  Secondly, the coherent evolution of the system potentially violates the directionality rules implied by all mancala games: there are projective measurement outcomes where ``stones'' move ``backwards'' in this quantum version. This phenomenon appears in certain outcomes shown in the last step of Figure~\ref{fig:board}(b).

Similar to other state engineering strategies, such as FUMES \cite{pedersen2014many} and the Zeno-locked FUMES (Z-FUMES) \cite{sorensen2018quantum}, ManQala uses two different methods to direct the evolution of a state: coherent unitary evolution and projective measurements. 
Unlike FUMES and Z-FUMES, ManQala starts by driving the system deterministically to a point where random search via projective measurements is more feasible. Once the deterministic strategy is complete, a stochastic approach, such as (Z-)FUMES, is adopted on one or more subsets of the board while the rest remains Zeno-locked.
Additionally, ManQala can also be implemented in a parallel fashion where a given lattice can be fragmented into sublattices, each of which is evolved independently.

As stated, the goal in ManQala is similar to Tchoukaillon or any other mancala variant; to end the game with all of the seeds in the Ruma. Thus, our aim in ManQala is to engineer a quantum state such that final state of the board mimics the end of a mancala game.  
While this goal appears restrictive, in the Appendix, we describe an approach for leveraging ManQala for quantum control problems with completely general target states.

\subsection{Formulation for systems with two-site bosonic hopping}

We now discuss the realization of ManQala-inspired QSE methods for controlling physical systems that exhibit two-site bosonic hopping due to their prevalence in practical situations.  
Bosonic hopping occurs in a variety of physical systems, including coupled optical cavities~\cite{nohama2007quantum}, coupled waveguides~\cite{ ogden2008dynamics}, transmons coupled to superconducting cavities~\cite{ma2021quantum}, and ultracold atoms in optical lattices  that obey the Bose-Hubbard (B-H) model~\cite{gersch1963quantum}. 
Implementation of the last two systems enable non-demolition measurement \cite{unruh1978analysis, braginsky1980quantum, belavkin1992nondemolition} of bosonic populations \cite{mekhov2009quantum, ma2021quantum,rogers2014characterization}. However, we restrict our states of interest to bosons on a lattice that evolve according to the one dimensional B-H model due to its simplicity in measurement \cite{rogers2014characterization}, and intuitive similarity to a traditional mancala game. 

Here, each lattice site that carries bosonic modes is analogous to a pit on a mancala board such that the quantum analogue of the sowing operation can be implemented by boson hopping. 
Therefore, the B-H Hamiltonian governing this scenario is given by,
\begin{gather}
    \hat{H}=-J\sum_{\langle i,j \rangle}{\hat{a}_i^\dagger\hat{a}_j}+\frac{V}{2}\sum_{i}\hat{n}_i(\hat{n}_i-1)-\mu\sum_i{\hat{n}_i}.
\end{gather}
Here, $\hat{a}_i^\dagger$ and $\hat{a}_i$ are bosonic creation and annihilation operators such that $\hat{n}_i=\hat{a}_i^\dagger\hat{a}_i$ gives the number of particles on site $i$ and $\langle i,j \rangle$ denotes summation over all neighboring sites $i,j$. The hopping amplitude $J$ describes the coupling strength between neighboring sites and the parameters $V$ and $\mu$ represent the on-site interaction and the chemical potential, respectively. 

By following the lead of S{\o}rensen et al.~\cite{sorensen2018quantum}, we restrict our method to quantum systems described solely by the quadratic hopping part of the B-H Hamiltonian as the self-interaction potential and the chemical potential are set to $V=0$ and $\mu=0$. Our motivation in doing so is two-folds. First, removing the terms that prevent tunneling aids the search for the target state, similar to using high temperature parameters in initial stages of simulated annealing algorithms \cite{aarts2005simulated}. Second, the quadratic B-H model is exactly diagonalizable in the Heisenberg picture resulting in closed-form, analytic solutions for unitary evolution once those terms are omitted \cite{rai2008transport, ogden2008dynamics}.

In general, the engineering of a quantum state is an optimization problem where the controller interacts with the given system via time-dependent actuation, $\vb{\alpha}=\left(\vb{\alpha}_0,\dots,\vb{\alpha}_t,\dots ,\vb{\alpha}_T\right)$, for times $t \in [0,T]$. 
Hence, the goal of ManQala is to specify $\alpha$ at all times using strategies derived from the game as shortcuts to optimization. 
In ManQala, the control actions at any given time $t$ are constrained to a set of unitaries and projective measurements, $\vb{\alpha}_t=\left(\vb{U}_t,\vb{P}_t\right)$. Here, a control action of pure coherent evolution, $\vb{\alpha}_{t_{ini}}=\left(\vb{U}_{t_{ini}},\vb{P}_{t_{ini}}=I\right)$,  implements the time evolution operator $\vb{U}_{t_{ini}}=\exp\left(-iH(t_{f}-t_{ini})\right)$ of the total Hamiltonian that drives the state from the initial time, $\ket{\psi(t_{ini})}$, to the final, $\ket{\psi(t_{f})}$, segment as the measurement projection operator is just the identity $I$. Similarly, an action of pure projective number state measurements $\vb{\alpha}_t=\left(\vb{U}_t=I,\vb{P}_t=\bigotimes_{j=0}^{M-1} P^{(j)}\right)$ has the effect of probabilistically collapsing the state. Here $\bigotimes_{j=0}^{M-1} P^{(j)}$ denotes the tensor product of the projective measurement operators at each site $j$, ranging from $0$ to $M-1$. And, combining those, the action $\vb{\alpha}_{t_{ini}}=\left(\vb{U}_{t_{ini}}=I,\vb{P}_{t_{ini}}= \bigotimes_{j\neq m} P^{(j)}\right)$ exerts coherent time evolution on all the sites except the sites at $j=m$ (for $m\in [0, M-1]$), unless those obscure inter-site tunneling, via Zeno-locked time-evolution, $U_{ZL}(t_{ini})=\vb{P}_{t_{ini}}\exp\left(-i\vb{P}_{t_{ini}}H\vb{P}_{t_{ini}}(t_{f}-t_{ini})\right)$ \cite{facchi2002quantum, facchi2008quantum,sorensen2018quantum}.

If, for the given Bose-Hubbard model, the task in hand is steering an initial state, $\ket{\psi_0}$, towards a target state, $\ket{\psi_{\text{targ}}}$, then the control problem can be formulated as a combinatorial optimization problem of finding the actions $\vb{\alpha}$ that would minimize a distance metric between states, $\alpha^{\star}=\underset{\vb{\alpha}}{\arg\min}\; d(\psi_0,\psi_{\text{targ}};\vb{\alpha})$ \cite{korte2011combinatorial}. Here this distance metric, $d$, to be minimized could be coming from Schr\"{o}dinger, as in the quantum fidelity $\mathcal{F}(\psi_1,\psi_2)=\bra{\psi_1}\ket{\psi_2}$, or Heisenberg picture dynamics for a series of actions or actuators implemented in time, $\vb{\alpha}$. Also, this metric can be combined with other goal functions within a cost function to be minimized, $\mathcal{L}=\sum_j \lambda_j d_j$. Here $\lambda_j$ denotes a Lagrange multiplier for a metric or goal $d_j$, and these multipliers can be chosen to make the cost function a convex sum, $\sum_j \lambda_j=1$.

For our system of interest, a 1-D bosonic lattice governed by a Bose-Hubbard Hamiltonian, we can define a distance metric based on the number of tunneling events. Pedersen et al. initially proposed such a metric for Fock states to judge the performance of the FUMES algorithm \cite{pedersen2014many}. We can generalize the Pedersen metric to any site population via the following compact form. Given the M dimensional particle number expectation value vectors $\vb{n}_A=\left(\langle n_0 \rangle_{\psi_A},\dots,\langle n_{M-1} \rangle_{\psi_A}\right)$, and, $\vb{n}_B$ for states (or density matrices) $\ket{\psi_A}$ and $\ket{\psi_B}$, the number of tunneling events is given by
\begin{align}
    d_{T}\left(\vb{n}_A,\vb{n}_B\right)&=\sum_{k=0}^{M-2}\left|\left(\vb{n}_A-\vb{n}_B\right)_k+\left(\vb{n}_A-\vb{n}_B\right)_{k+1}\right|.
\end{align}

The global optimum of the number of tunneling events and the fidelity are the same when steering a system from an initial state to a target state. The same cannot be said for the intermediate steps. If we are given specific initial and target states, or their particle number expectation value vectors ($\vb{n}_0$ and $\vb{n}_\text{targ}$, respectively), we can define a bosonic distance akin to quantum (in-)fidelity for the particle number expectation value vector, $\vb{n}_A$, of a (probably unknown) state, $\ket{\psi_A}$, using the number of tunneling events given above, as the following,
\begin{align}\label{eq:bosonic_dist}
    d_B\left(\vb{n}_A,\vb{n}_{\text{targ}};\vb{n}_0\right)&=1-\frac{d_{T}(\vb{n}_A,\vb{n}_{\text{targ}})}{d_{T}(\vb{n}_0,\vb{n}_{\text{targ}})}.
    \end{align}
In this current form, the bosonic distance metric starts from zero for $\vb{n}_A=\vb{n}_0$ and takes on the maximum value of unity when $\vb{n}_A=\vb{n}_\text{targ}$ due to scaling with the number of tunneling events between the initial and target states. 

ManQala initially tries to minimize a cost function
\begin{equation}\label{eq: loss}
\mathcal{L}_M=\lambda_1 \left(1-d_B\left(\vb{n}(\vb{\alpha}),\vb{n}_{\text{targ}};\vb{n}_0\right)\right)+\lambda_2 N_P(\vb{\alpha}) + \lambda_3 M_C(\vb{\alpha}).
\end{equation}
Here the term $\vb{n}(\vb{\alpha})$ denotes site populations after the actions $\alpha$ were implemented on the initial populations and $N_P(\vb{\alpha})$ denotes the number of projective measurements we apply to our system as actions.
Assuming a high $\lambda_2$, we try to avoid stochastic methods (e.g, Z-FUMES) while minimizing $d_B$, but use deterministic unitary actions (site population permutations) instead. Also, if we want to be consistent with mancala, we can penalize the actions that are inconsistent with it via a term $M_C(\vb{\alpha})$. For example, we can penalize the unitary actions that do not follow the mancala's sowing rule. On the other hand, this condition can be relaxed by tuning $\lambda_3$, or dropped altogether ($\lambda_3=0$) as a modified-ManQala (mod-ManQala) algorithm without losing any generality, while converging faster.  This is considered in appendix ~\ref{sec:examples} and can be achieved by Zeno-locking all other sites.
 
Of note, a target board to be reached (e.g, winning condition) can be decomposed into sub-boards where each sub-board population can be thought of as a pit. For example, for the three pit board configuration we examine here, the target board can be decomposed into sub-boards (sub-lattices) of $\{2-pits,1-pit\}$. For such segmentation of the board, the target board is demarcated into sub-board populations $\left(3,0,0\right)\rightarrow((3,0),0)\rightarrow \left(3,0\right)$. Using that, we can also split our initial board configuration in hand the same way $\left(0,1,2\right)\rightarrow ((0,1),2)\rightarrow \left(1,2\right)$. Both in the traditional and the quantum game, bringing the board into the winning condition first requires bringing it to the target sub-board ($\left(3,0\right)$ in our example). The winnable board configurations enable the player to move the stones from the pits with small number of stones right away into the target sub-board, effectively decoupling from the move of the pits with larger number of stones that could overshoot the Ruma.  In ManQala, whether or not mimicking each traditional move, this winnable board intuition is actualized by moving the particles from the sites with a few number of particles right away to their designated smaller sub-lattices deterministically only via unitary evolution so that we can Zeno-lock them. This way we can move the particles from sites with many particles to their designated, larger, target sub-lattice. Having locked the (sub)-lattices with few particles, and moved the rest to their sub-lattice, we can use probabilistic search methods via measurement back-action on this previously unlocked sublattice to find the right set of actions that would steer our system to the target state. 

\begin{table}[t]
\centering
 \begin{tabular}{|p{2cm}|p{1.5cm}|p{0.6cm}|p{0.6cm}|p{1.2cm}|p{0.5cm}|} 
 \hline
 Strategy & Zeno-lock & $\mathcal{F}$ & $d_B$ & Parallel & $M_C$\\ [1ex] 
 \hline\hline
 ManQala & \checkmark & \checkmark &  \checkmark &  \checkmark & \checkmark \\ 
 \hline
 modified-ManQala &  \checkmark &  \checkmark &  \checkmark &  \checkmark & $\times$\\
 \hline
 FUMES & $\times$ &  \checkmark & $\times$ & $\times$& $\times$ \\
 \hline
 Z-FUMES &  \checkmark &  \checkmark &  $\times$ & $\times$& $\times$ \\ [1ex] 
 \hline
 \end{tabular}
\caption{\textbf{Summary of strategies compared in this paper.} FUMES conducts a greedy, stochastic, non-parallelizable search based on fidelity $\mathcal{F}$ alone. Z-FUMES conducts the same search, but gradually shrinks the search-space via Zeno-lock. Both (mod-)ManQala start by minimizing bosonic distance, $d_B$, deterministically via permuting the populations of one two or three-site at a time 
 to minimize stochasticity \& shrink search space and then implements Z-FUMES in parallel at various sub-lattices. Unlike ManQala, mod-ManQala does not necessarily adhere to the rules of mancala, as expressed by term $M_C$ in Eq.~\ref{eq: loss}. }
\label{table:compare}
\end{table}

To help clarify the differences between ManQala, FUMES, and Z-FUMES we include Table \ref{table:compare}.
A more detailed description of these differences in available in Appendix \ref{app:compare}.

\section{Numerical Simulations and Performance Comparisons}
\label{sec:two_site}

In the previous section, we presented the theoretical framework for a game-inspired quantum state engineering (QSE) strategy we call ManQala. 
Here, we provide numerical simulations of the performance of ManQala and compare them against FUMES and Z-FUMES.
For clarity, we will consider in this section the same problem illustrated in Figure \ref{fig:board}(b). 
Here, the initial state is given by $\ket{\psi}_0=\ket{0,1,2}_{\text{Fock}}$ and the target state by $\ket{\psi}_t=\ket{3,0,0}_{\text{Fock}}$ where each ket represents, for example, modes in a bosonic lattice. 
 The Hilbert space dimension is $R=\binom{N+M-1}{N}=10$ for $M= 3$ sites and $N=3$ particles.  Thus, in this configuration, there are $R=10$ possible Fock states. 

Assuming prior knowledge of the total number of particles $N$ in the system, we can deduce the state of the entire lattice by measuring only $M-1$ of the sites. 
Hence, for the problem we consider here with $N=M=3$, we can represent the state of the system as a sequence of two numbers, in this case, the number of particles in the leftmost site (Ruma) and the next nearest site.  We denote all ten possible configurations as
$\vb{L}=[\vb{l}^{(0)}\mathbin{,}\dots\mathbin{,}\vb{l}^{(9)}]=[(0\mathbin{,}1)\mathbin{,}(0\mathbin{,}0)\mathbin{,}(0\mathbin{,}2)\mathbin{,}(1\mathbin{,}0)\mathbin{,}(1\mathbin{,}2)\mathbin{,}(2\mathbin{,}1)\mathbin{,}(1\mathbin{,}1)\mathbin{,}(0\mathbin{,}3)\mathbin{,}(2\mathbin{,}0)\mathbin{,}(3\mathbin{,}0)]_{\text{Fock}}$. 
In the particular example shown in Figure \ref{fig:board}(b), $\vb{l}^{(0)}$ and $\vb{l}^{(9)}$ are the initial and target states respectively. 

The FUMES protocol uses unitary evolution to maximize the fidelity with the final state followed by a single projective measurement. 
To find these unitary evolutions, we first identify the distance between all possible states $\vb{L}$ and the target state. For each state in $\vb{L}$, there is a corresponding unitary time-evolution duration that maximizes the probability of observing the target state and can be also written in a vector form.
We write these times again in a vector of the form $\vb{T}=\left[t_{\text{desig}}^{(0)},\dots,t_{\text{desig}}^{(9)}\right]$. 
For the target state $\vb{l}^{(9)}$ we obtain these values using QuTiP ~\cite{johansson2012qutip} by numerically solving Schr\"{o}dinger's equation via exact diagonalization. The resulting time scales are given by $\vb{T}=[1.66\mathbin{,}2.22\mathbin{,}1.33\mathbin{,}1.33\mathbin{,}0.89\mathbin{,}0.555\mathbin{,}1.11\mathbin{,}0.11\mathbin{,}0.89\mathbin{,}0]$. 

If we instead consider situations where we Zeno-lock certain sites, such as in Z-FUMES, the optimal time-evolution values change.  Zeno-locking only occurs in this example when the $j=2$ site has the targeted number of particles (in this case zero particles) in it, $\ket{\cdot, \cdot, 0}$, and not the others, $\ket{\cdot, 0, \cdot}$ and $\ket{0,\cdot,\cdot}$. Locking the $j=0$ (Ruma) site when it has zero particles would be counterproductive as the target state has nonzero particles in that site. However, one may be tempted to lock the intermediate site ($j=1$) when it has the target number of particles (zero), but this would obstruct the tunneling between the sites $j=0$ and $j=2$. 
Zeno-locking the site $j=2$ when it has zero particles in it reduces the set of possible configurations $\vb{L}$ to the subset $\vb{L}_Z=[(1\mathbin{,}2)\mathbin{,}(2\mathbin{,}1)\mathbin{,}(0\mathbin{,}3)\mathbin{,}(3\mathbin{,}0)]_{\text{Fock}}$. 
 Designated time durations become the following for this case, $\vb{T}_Z=[0.953\mathbin{,}0.615\mathbin{,}1.567\mathbin{,}0]$, where the designated time duration vector for the states to be evolved while Zeno-locking is replaced with the ones from the usual FUMES while keeping the rest same.  

 Hilbert space reduction using Zeno-locking, as shown in the difference in the dimensionality between $\vb{T}$ and $\vb{T}_{Z}$, is precisely why Z-FUMES outperforms FUMES in general. 
 Note, however, that Z-FUMES only applies Zeno locking in a stochastic fashion. 
For example, if Z-FUMES encounters a $0$ on the second site and Zeno-locks it so that the two-site sublattice to the left has the correct amount of particles, the search space reduces to $R=\binom{N+M-1}{N}=4$ for ($N=3, M=2$). 
In comparison, ManQala and its variants drive the system deterministically exactly to this configuration instead of waiting to encounter it during the random search.
Finally, we note that since FUMES does not use Zeno-locking at all the algorithm searches the entire $R=10$ dimensional space. Hence, intuitively, we expect both ManQala and Z-FUMES to outperform FUMES in terms of resources required to reach the target state.

To reach the target state using ManQala, that is limiting the ability to permute to two-sites at a time while adhering to the sowing rules of mancala, we begin by driving our system to the $\vb{L}_{Z}$ subspace and Zeno-locking the $j=2$ mode. To achieve this we use three two-site unitaries of $\Delta t=\pi/2$. Here each unitary permutes the particle population of adjacent Fock states of sites $j$ and $j+1$ carrying $N$ particles, $U(\pi/2)\ket{k, N-k} = \exp\left(-i(a_ja_{j+1} + \text{H.C})\pi/2\right)\ket{k,N-k}= \ket{N-k,k}$. In the case given in Figure \ref{fig:board}, we need a total deterministic evolution time of $\Delta t = 3\pi/2$ to drive the system into the Zeno-locked configuration of $\ket{2,1,0}$. Once there we apply the aforementioned Z-FUMES with time durations $T_Z$.
 
Alternatively, if we instead use mod-ManQala, we require only one three-site unitary of duration $\Delta t = \sqrt{2}\pi/2$ to deterministically reach the Zeno-locked configuration due to relaxing the need to imitate mancala. Here each unitary acting on Fock states of sites $j, j+1, j+2$ permutes the populations of sites $j$ and $j+2$, while leaving $j+1$ untouched. The three-site unitary  given by $U(\sqrt{2}\pi/2) = \exp\left(-i\sqrt{2}(a_ja_{j+1} + a_{j+1}a_{j+2} + \text{H.C})\pi/2\right)$ implements the permutation $U(\sqrt{2}\pi/2)\ket{k,l,N-l-k}= \ket{N-l-k,l,k}$.  To be more specific, consider a demarcation of the three-dimensional lattice in sub-lattices $\left\{2-site, 1-site\right\}$, i.e, $ \vb{\tilde{S}}^{\star}=\left[s_1,s_2\right]=\left[(0,1),2\right]$. 
The initial site population for this $\vb{\tilde{S}}^{\star}$ sub-lattice segmentation is $\vb{n}_0=\left((0,1),2\right)=\left(1,2\right)$. 
ManQala tries to be as consistent as possible with the winning strategy of the analogous classical Tchoukaillon board and hence drives the system into $\vb{n}_M^{\star} = \left((2,1),0\right) = \left(3,0\right)
$ populations with respect to segmentation $\vb{\tilde{S}}$ and corresponding permutation $\tilde{\pi}$. The permutation to drive the initial populations into the Zeno-locked sub-lattice $\left(3,0\right)$ is the following: 
\begin{equation}
\tilde{\pi}_M^{\star}=\begin{pmatrix}
0 & 1  & 2  \\
2 & 1  & 0 
\end{pmatrix}
\end{equation}

Here $\tilde{\pi}_{M^{\star}}$ denote the site population permutations for both the ManQala and mod-ManQala strategies, i.e, exchanging sites $0$ and $2$ while keeping site $1$ the same.
Its corresponding matrix representations $\tilde{P}_{\tilde{\pi}_{M^{\star}}}$ is a column-reversed identity matrix.
In the segmentation $\vb{\tilde{S}}^{\star}$, both of these representations yield $\eta=\left(\tilde{P}_{\tilde{\pi}}\vb{n}_0-\vb{n}_{\text{targ}}\right)=\left(0,0\right)$. 
In practice, we first identify these permutation operators and then compile them into two and three-site unitaries (i.e, ``moves" in the previous section) based on the traditional game constraints or lack thereof.  

\begin{figure}[t]
\centering
\includegraphics[width=\linewidth]{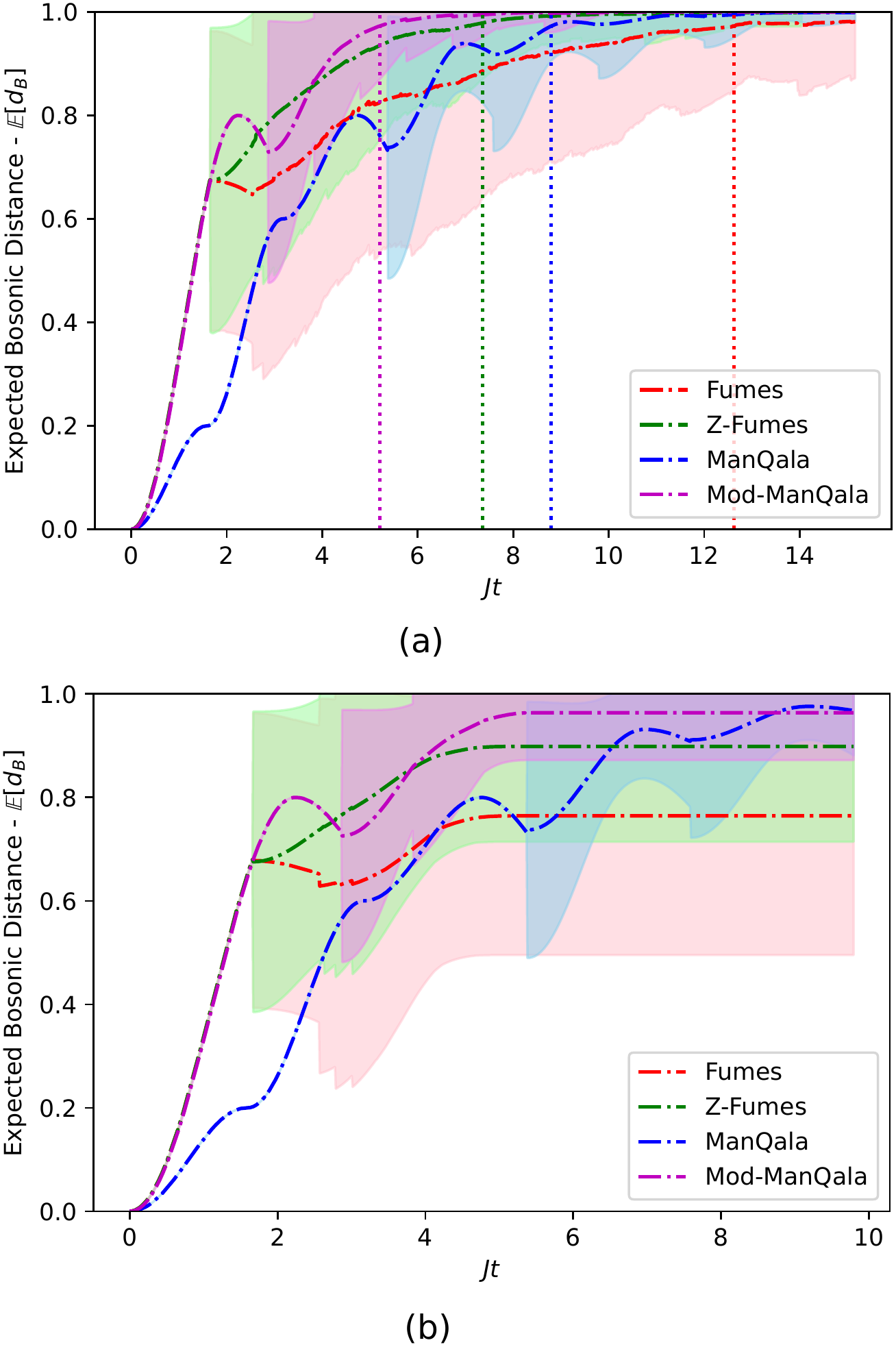}
 \caption{\textbf{Numerical simulation of game-inspired quantum state engineering}. 
 Given the initial state $\ket{0,1,2}$, the average (expected) bosonic distance of observing the target state $\ket{3,0,0}$ for different quantum state engineering strategies computed using $1000$ stochastic trajectories each in QuTiP. (a) For each stochastic trajectory the non-deterministic part of a given strategy is repeated until convergence. The dashed lines mark when each strategy achieves $0.99$ bosonic distance, $d_B$ (defined in the text). (b) Non-deterministic part is only repeated twice for each stochastic trajectory. Red (FUMES), green (Z-FUMES), blue (ManQala), and purple (mod-ManQala) curves denote the average $d_B$ of a given strategy over $1000$ trajectories, while the shaded areas are the respective standard deviations. Note that a projective measurement does not occur until the emergence of a shaded area (coloring) corresponding to a standard deviation.}
 \label{fig:compare1}
\end{figure}

Figure~\ref{fig:compare1} shows the comparison of expected bosonic distance between different quantum state engineering strategies for initial state $\ket{\psi_0}=\ket{0,1,2}$ and target state $\ket{\psi_{\text{targ}}}=\ket{3,0,0}$ over a thousand stochastic, Monte-Carlo trajectories. In terms of the bosonic distance, FUMES, Z-FUMES, and mod-ManQala start with a steep linear ramp as they all implement a three-site interaction. When the fidelity between the state in hand and the target state is maximized at $J t_{\text{desig}}^{(0)} = 1.66$ both (Z-)FUMES apply a projective measurement, leading to different stochastic trajectories (green and red coloring in the Figure \ref{fig:compare1} representing the standard deviation of these). Both (mod-)ManQala aim to achieve $\ket{2,1,0}$ deterministically, and their bosonic distances ascend during the course of their time-evolution, albeit slowly for ManQala due to constraints. Once ManQala achieves a local peak at $Jt=3\pi/2$ and mod-ManQala achieves the same at $Jt=\sqrt{2}\pi/2$, they make a downward turn as they are both optimizing for fidelity now (using Z-FUMES) by both evolving for $Jt_{\text{desig}}^{Z}=0.615$ in the Zeno-locked space (i.e, the second element of $\vb{T}_Z$), and make projective measurements at those spots (leading to blue and purple coloring for their respective variances). 

In this Zeno-locked space mod-ManQala follows the designated measurement times $\vb{T}_Z$ based on measurement results, while ManQala, following the rules of the traditional game, always brings the system back to the $\ket{2,1,0}$ and repeats. Because of this, the mean (expected) trajectory progresses differently in time for ManQala, and mod-ManQala converges to a $d_B = 0.99$ much faster ($Jt \sim 5.1$), although they achieve the same exact statistics given in part $b)$ of Figure \ref{fig:compare1}. 
Since we are driving the system into Zeno-locked subspaces determinisitically, and querying a smaller search space (mod)-ManQala has much less variance. In Figure~\ref{fig:compare1}(a) each protocol is repeated indefinitely until achieving a unity fidelity such that FUMES has an average standard deviation of $0.21$, while Z-FUMES has $0.17$, mod-ManQala has $0.06$, and ManQala has $0.06$ in bosonic distance until observing $0.99$ bosonic distance. This phenomenon is much more pronounced in  Figure~\ref{fig:compare1}(b)
as just a small number of protocol repetitions is enough to achieve near unity fidelities with the target state. 
For completeness, we provide results related to the repeated application of all three protocols in Appendix \ref{app:rep}.

\begin{figure}[t]
    \centering
    \includegraphics[width=\linewidth]{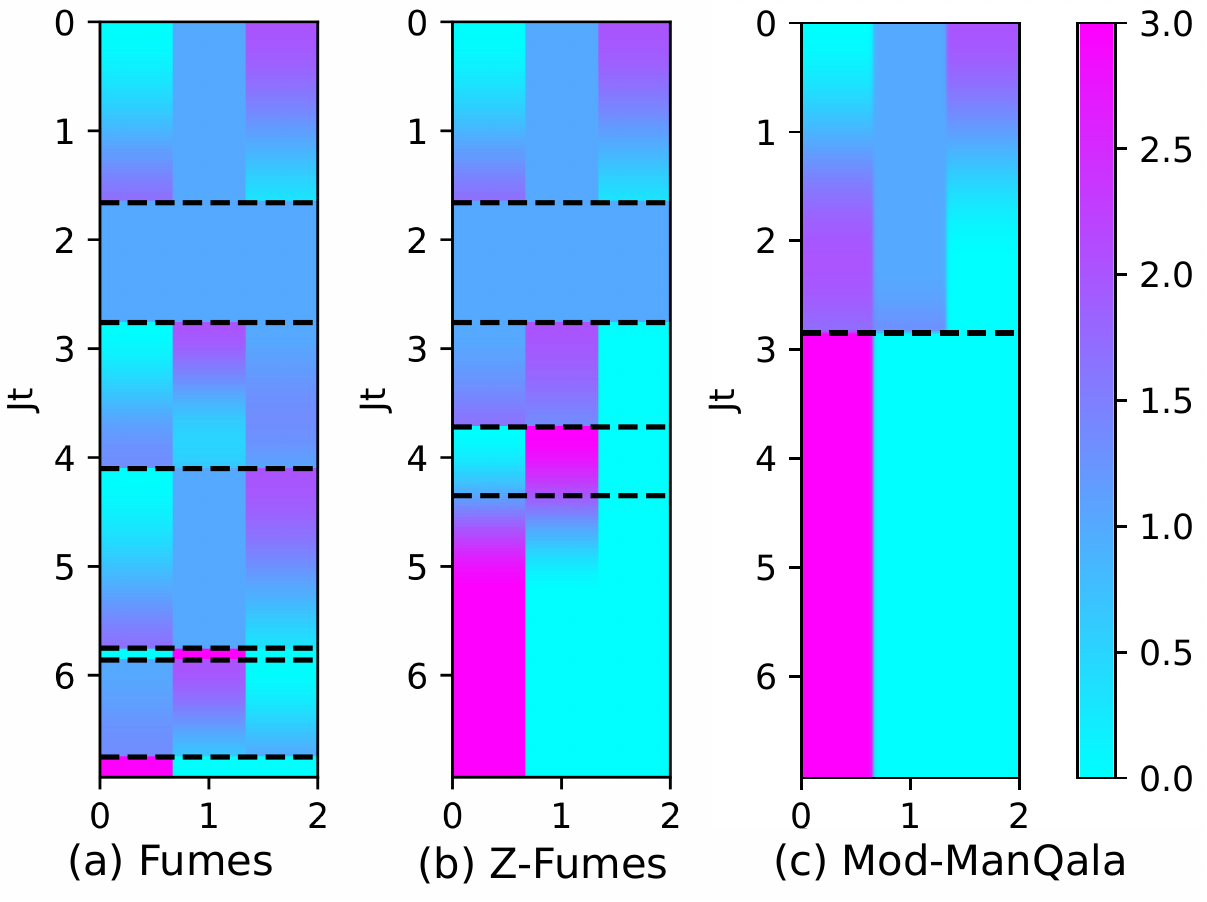}
    \caption{\textbf{Time evolutions of the expected number of particles for three selected stochastic trajectories:  (a) FUMES, (b) Z-FUMES and (c) mod-ManQala}. The horizontal axis shows the site number and the color corresponds to the expected particle number. The dashed lines represent the times a projective measurement occurs. Note in (a) and (b), the time between the first and second projective measurements is where the Heisenberg picture dynamics halt (Mott-insulator state). In (c) there is only one probabilistic event (measurement). This Figure represents a single (randomly chosen but representative) stochastic instance for each strategy out of many possibilities.}
    \label{fig:compare2}
\end{figure}

To further illustrate the difference between these quantum-state engineering strategies, we compare the expected particle numbers at each site with the same aforementioned initial and target states, as shown in Figure ~\ref{fig:compare2}. This figure shows individual (randomly chosen but representative) evolutions for FUMES, Z-FUMES and mod-ManQala as compared to Figure \ref{fig:compare1} which averages many such evolutions to find averages. While not as useful to understanding the general behavior of each protocol, Figure \ref{fig:compare2} provides an intuitive understanding of how (Z-)FUMES and mod-MaQala strategies differ in their overall approach to solving the same problem. Note that the dashed lines in Figure \ref{fig:compare2} indicate projective measurements. (Z-)FUMES implements a measurement each time the probability of observing the target state is maximized in a given (Zeno-locked) configuration. Alternatively (mod-)ManQala strategies evolve the sites deterministically towards states where Zeno-locking can be used to reduce the Hilbert space (in this case making the population in the $j=2$ mode $0$). 
As pictured in Figure \ref{fig:compare2}, this particular stochastic run is successful at projecting to the correct state in panel (c) on the first projection, however, if this were not the case mod-ManQala would again try to maximize fidelity before repeating the projective measurement. Note that (Z-)FUMES evolves the states in time for the next projective measurement even when the particle expectation numbers, as well as the bosonic distance we defined $d_B$, are frozen in time. In parts a) and b) of Figure \ref{fig:compare2}, the first measurement represented by a dashed line projects the system into a Mott-Insulator state $\ket{1,1,1}$, where particle expectations are not changing in time under the effect of unitary evolution. The second projective measurement, represented by a second dashed line, takes them out of this state.

In this section, we have considered scenarios where the target state mimics the end state of the original solitaire mancala game, meaning all bosons end up in a single mode. However, we note that ManQala can be applied more generally to problems with arbitrary target states. These general approaches require us to loosen our adherence to  mancala (game rules, directions). For concreteness, in Appendix \ref{sec:examples} we describe how ManQala can be applied to two important physical systems, those of superfluids and Mott-insulator systems with site and particle numbers of five. 
In particular, we show in Appendix \ref{sec:examples} how the ManQala framework can divide more significant, hard-to-tackle problems into small, manageable, and parallelizable ones.

\section{Conclusion}
\label{sec:conclusion}
In this paper, we devised a quantum engineering strategy that we term ManQala, inspired by the traditional solitaire game Tchoukaillon and illustrated the differences between our approach and other competing strategies. In particular, we provided numerical comparisons of ManQala and ManQala-inspired strategies against FUMES and Z-FUMES. 
In all cases we found that ManQala strategies ultimately match or outperform FUMES and Z-FUMES.

ManQala augments existing quantum state engineering strategies by adding a preprocessing stage that consists of deterministic unitary permutations. These permutations reduce the Hilbert space of the problem, improving the performance of search-based strategies such as FUMES and Z-FUMES. More specifically, FUMES is a greedy, stochastic algorithm that optimizes the fidelity between the initial and target state through unitary evolution of the whole Hamiltonian followed by the collapse of the state by observation. On the other hand, Z-FUMES uses the same stochastic and greedy algorithm as FUMES but with the additional feature that if we end up in subspaces of the target state during probabilistic jumps, we can Zeno-lock these subspaces and only evolve the remaining sites. By contrast, ManQala-based strategies intentionally drive the system into target subspaces/sublattices of the overall state through deterministic unitary evolution into configurations that allow for an overall reduction in Hilbert space size through the Zeno-locking of certain modes. Then, once demarcated, ManQala strategies use local algorithms to control each subspace/sublattice to minimize the bosonic hopping distance. ManQala's use of subspaces and sublattices naturally lends itself to parallelization, which would further improve the performance of ManQala over existing strategies as problem spaces increase in dimension. 

In our formulation of ManQala, we have chosen to use the bosonic hopping distance to inform the algorithm's actions instead of overall fidelity. Since ManQala focuses on sublattices, within the execution of the protocol, the overall fidelity may decrease before rapid improvement. This decrease in fidelity is due to the algorithm focusing on bosonic distances between sublattices at any individual step and not the global fidelity. Ultimately our quantum version of a mancala game has provided a helpful framework for thinking about state engineering in quantum systems. The observed performance advantages of ManQala over competing strategies suggest an exciting connection between  games and quantum systems engineering, and future work should continue exploring this relationship.

\begin{acknowledgments}
We acknowledge the primary support of this work from the IBM-HBCU Quantum Center at Howard University.  Additionally, the views and conclusions contained in this document are those of the authors and should not be interpreted as representing the official policies, either expressed or implied, of the Army Research Laboratory or the U.S. Government. The U.S. Government is authorized to reproduce and distribute reprints for Government purposes notwithstanding any copyright notation herein. Additionally, this material is based upon work supported by, or in part by, the Army Research Laboratory and the Army Research Office under contract/grant numbers W911NF-19-2-0087 and W911NF-20-2-0168.  TAS's contribution is based upon work supported by the U.S. Department of Energy, Office of Science, National Quantum Information Science Research Centers, Co-design Center for Quantum Advantage (C2QA) under contract number DE-SC0012704.
\end{acknowledgments}

\bibliography{apssamp}

\appendix

\section{Algorithm}
\begin{algorithm}[H]
\caption{(mod-)ManQala}\label{alg:m_manqala_x}
\SetKwInOut{Input}{Input}
\SetKwInOut{Require}{Require}

\Input{Initial state $\rho_0$, Target state $\rho_{\text{targ}}$, Total Hamiltonian $H$,A Boolean for choosing algorithm \textit{M\_ManQala}}
\Require{Full knowledge of $\rho_0$, Target state being a multi or single site Fock state $\rho_{\text{targ}}$}
\begin{algorithmic}
\State $\vb{n}_{\text{targ}}=$\textbf{n\_ExpectationValues}$(\rho_{\text{targ}})$\\
\State \textbf{Initialize: } $\rho_t=\rho_0$\\
\If{$\rho_t$ \textbf{is} \textit{SuperPositionState}}{
\State $\rho_t, \vb{n}_t=$\textbf{MeasureAllSites}$(\rho_t)$
\Comment{Get the new collapsed (Fock) state, and its corresponding measurement sequence}
}
\State $\Tilde{\vb{s}},\Tilde{\vb{\pi}}=$\textbf{DemarcateSubLattices}$(\vb{n}_{\text{targ}},\vb{n}_t)$ \Comment{Outputs a list of designated sub-lattices, and the permutations needed to bring the state's sub-lattice populations into that of target's.}\\
\Comment{e.g, $\Tilde{\vb{s}}=(2,1)$ if 3-site lattice is demarcated into a sub-lattice of 2 and 1}\\
\Comment{e.g, $\Tilde{\vb{\pi}}$ is a 2 x M dimensional matrix showing the permutations of ManQala moves}\\
\Comment{e.g, In the paper, sites 0 and 1 are switched, then 2 and 1 permuted}\\
\State $\vb{n}\_{\text{targ}}\_sl=$\textbf{Get\_n\_SL}$(\vb{n}_{\text{targ}},\Tilde{\vb{s}})$ \Comment{Sub-lattice populations of the target state given demarcations.}
\State $\vb{n}\_t\_sl=$\textbf{Get\_n\_SL}$(\vb{n}_t,\Tilde{\vb{s}})$ \Comment{Sub-lattice populations of the initial state given demarcations.}
\State $\vb{P}_{list},t_{list}=$\textbf{ManQalaMoves}$(\Tilde{\vb{s}},\Tilde{\vb{\pi}},$\textit{M\_ManQala}$)$\\
\Comment{$t_{list}$ is the time durations of unitary evolution operators doing permutations.}\\
\Comment{$\vb{P}_{\text{list}}$ is a list of projective operators to be Zeno-locked. They are just identities when the Zeno-locking condition is not met.}
\State $d_B=$\textbf{BosonDistance}$(\vb{n}\_t\_sl,\vb{n}_{\text{targ}}\_sl)$
\Comment{Bosonic distance between the state's, and target state's sub-lattice populations.}
\State $\vb{\alpha}=[]$
\State \textit{counter}$=0$\\
\While{$d_B < 1$}{
\State $t,P=\vb{P}_{\text{list}}$[counter],$t_{\text{list}}$[counter] \Comment{Get two or three-site unitary moves to do site population permutations}
\State $U=P\exp\left(-itPHP\right)$
\State $\vb{\alpha}$.\textbf{append}$(U)$
\State $\rho_t=U\rho_tU^{\dag}$
\State $\vb{n}_{t}=$\textbf{n\_ExpectationValues}$(\rho_t)$
\State $\vb{n}\_t\_sl=$\textbf{Get\_n\_SL}$(\vb{n}_t,\Tilde{\vb{s}})$
\State $d_B=$\textbf{BosonDistance}$(\vb{n}\_t\_sl,\vb{n}\_{\text{targ}}\_sl)$
\State \textit{counter}$=$\textit{counter}$+1$
}
\State $\vb{\alpha}$.\textbf{append}(\textbf{Z\_FUMES}$(\rho_t,\rho_{\text{targ}},H)$)
\end{algorithmic}
\SetKwInOut{Output}{Output}
\Output{$\vb{\alpha}$}
\Comment{Set of unitary actions from ManQala, then unitary actions and projections measurements from the final Z-FUMES}
\end{algorithm}

\section{Details of the comparison between ManQala, FUMES, and Z-FUMES}
\label{app:compare}

We now describe the general differences between the ManQala strategy and the FUMES and Z-FUMES strategies. The FUMES strategy unitarily evolves an initial quantum state to the point where the probability of observing the target quantum state is maximized, then performs projective measurement in an attempt to collapse the system to the desired target \cite{pedersen2014many}. The experimenter stores the classical information of the measurement read-out (e.g, in a classical register), and checks whether the observed measurement sequence $\vb{l}^{(r)}$, a member of all the $R$ possible measurement sequences $\vb{L}=[\vb{l}^{(0)},\hdots,\vb{l}^{(R-1)}]$, matches the one of the target state. If the said projective measurement does not collapse the state onto the target state, then the process is repeated by evolving the state via unitary evolution for a designated duration, $t_{\text{desig}}^{(r)}$, based upon the measurement results until convergence \cite{pedersen2014many}. These durations are computed ahead of the experiment by solving the deterministic Schr\"{o}dinger equation for each possible intermediate state, upon which system could collapse onto, with respect to the target state \cite{pedersen2014many}. The FUMES strategy is indirect and always tries to go to the nearest, optimal solution.

The Z-FUMES strategy makes use of the Zeno subspaces to demarcate the bosonic lattice into sub-lattices. If the sublattices and bosonic sites have the targeted number of particles, they are Zeno locked from tunneling into adjacent sites through frequent application of projective, non-demolition, particle number measurements \cite{sorensen2018quantum,facchi2002quantum,facchi2008quantum}. Just like FUMES, there is a designated time-evolution duration, $t_{\text{desig}}^{(r)}$, for each particle number measurement sequence $\vb{l}^{(r)}$  of $r=0\hdots R-1$ within the set of possible measurement sequences $\vb{L}=\left[\vb{l}^{(0)},\hdots\vb{l}^{(R-1)}\right]$. On the other hand, some of the states (or their corresponding sequences) $\vb{l}_Z^{(r)}$ within the $\vb{L}$ warrants Zeno locking of specific sites, then evolving the lattice for $t_Z^{(r)}$. Conversely, if a sequence $\vb{l}_{NZ}^{(r)}$, which does not require Zeno locking any sites, is observed, usual FUMES is implemented for the designated time duration $t_{NZ}^{(r)}$. FUMES is a greedy algorithm as it chooses the ``shortest visible path" in minimizing a cost function $\mathcal{L}_{\text{FUMES}}(t)=\mathcal{F}(\psi(t),\psi_{\text{targ}};\vb{\alpha})$ at given time $t$. Z-FUMES is also a greedy algorithm, but it puts a Zeno-locking constraint on the actions $\alpha$ based on the site-population measurement outcome, $\vb{n}(t^{-})$, right before the implementation at $t$. That Zeno-locking constraint can be represented in the Z-FUMES cost function via a Lagrange multiplier,   $\mathcal{L}_{\text{Z-FUMES}}(t)=(1-\lambda)\mathcal{F}(\psi(t),\psi_{\text{targ}};\vb{\alpha}) + \lambda Z(\alpha,\vb{n}(t^{-}))$ at given time $t$.  Here $Z$ denotes the Zeno-locking constraint on the unitary actions to be taken if the outcome of the projective measurements finds sub-lattice populations that satisfy Zeno-locking condition (i.e, having target state populations, and no tunneling boundary).  \linebreak

Both ManQala and modified-ManQala strategies begin by optimizing $\mathcal{L}_M$ via unitary actions. However, in modified-ManQala, those two or three-site unitary actions do not necessarily need to follow each  mancala move (i.e, $\lambda_3 = 0$). Once they reach a plateau where $d_B$ can no longer be improved via those unitary actions (permutations), i.e, a plateau where we can Zeno-lock some site populations, both of these strategies employ Z-FUMES in parallel, or in sequence at different sub-lattices. In other words, they optimize multiple $\mathcal{L}_{\text{Z-FUMES}}(t)$. These site population permutations are implemented through two-site unitary time evolutions with duration $\pi/2$ and the three-site ones with duration $\sqrt{2}\pi/2$ (again via selective Zeno-locking). We drive the system into Zeno-locked sub-lattices through those permutations implemented by unitary evolution. We restrict the strategy to 2 \& 3 site interactions because they yield site population permutations for characteristic time durations. Thus, the problem of finding permutations, $\Tilde{\vb{\pi}}$, between site populations, and demarcating the lattice sites $[0,M-1]$ into $m$ sub-lattices $\vb{\tilde{S}}=[s_1,\dots,s_j,\dots s_m]$ to be Zeno-locked are equivalent problems and tackled at the same time. Given the initial state and target state populations, $\vb{n}_0$ and $\vb{n}_{\text{targ}}$, the problem can be reduced into classically solving the following combinatorial optimization problem before applying any controls,

\begin{align}
    \tilde{\vb{S}}^{\star}, \tilde{\vb{\pi}}^{\star} = \underset{\tilde{S},\tilde{\pi}}{\arg\min} \sum_{s_j}^m\left|\sum_{i_{s_j}\in s_j}^{m_j}\left(\tilde{P}_{\tilde{\pi}}\vb{n}_0-\vb{n}_{\text{targ}}\right)_{i_{s_j}}\right|.
\end{align}

If we get the answer for this optimization problem, we would acquire sub-lattices $\tilde{\vb{S}}^{\star}$, and permutations $\tilde{\vb{\pi}}^{\star}$ that will be decomposed into designated unitary operators with time durations. Here $\tilde{\vb{\pi}}$ denotes a permutation (group), $\tilde{P}_{\tilde{\pi}}$ denotes the matrix representation of the said permutation (group), and super-scripts $(\dots)^{\dots}$ denote the solutions found \cite{sternberg1995group}. We want to find permutations $\tilde{\vb{\pi}}$ such that we could split $[0,M-1]$ into $m$ different sets of adjacent sites $\vb{\tilde{S}}=[s_1,\dots,s_j,\dots s_m]$ based on a condition on the vector $\eta=\left(\tilde{P}_{\tilde{\pi}}\vb{n}_0-\vb{n}_{\text{targ}}\right)$. We could split it, if and only if, sum of the $m_j$ elements of each set $s_j$ is equal to zero. The trivial example is when we have $m=1$ sub-lattices such that $\vb{\tilde{S}}=[s_1]=[0,\dots,M-1]$, and the corresponding trivial representation is $\tilde{P}_{\tilde{\pi}}=I_M$. Here $I_M$ denotes the $M$ dimensional identity matrix corresponding to the matrix representation of the permutation, $\tilde{\pi}$, of site populations. The general solution to this problem is marked with the \textbf{DemarcateSubLattices} function in the ManQala pseudo-code (see Alg. \ref{alg:m_manqala_x}, also in the Appendix). Similarly, we marked the task of compiling these permutations into designated unitaries with specific time durations with the function \textbf{ManQalaMoves}.

In Z-FUMES, we passively search for these Zeno-locked sublattices to narrow our search space while randomly searching for the target state actions. However, those are random encounters. In ManQala we designate the target sub-lattices to demarcate the lattice into, then we actively and deterministically drive our system to them via two and/or three site unitary evolution. To do so, we search for these unitary moves that take less time while minimizing the bosonic distance $d_b$ between the particle expectations of the sub-lattices of the state in hand, and the ones of the target state. Once we drive them, we can Zeno-lock and/or run a Z-FUMES probabilistic search on these sub-lattices. 
Note, in the main text we consider problems with $N=M=3$, which is the smallest scale problem where Zeno locking can be applied. Even at this small scale an observable difference in performance between ManQala and (Z-)FUMES, which we expect to only increase for larger problems where shrinking the search space is more important.

\section{Repetitions}
\label{app:rep}

\begin{figure}[ht]
    \centering
    \includegraphics[width=\linewidth]{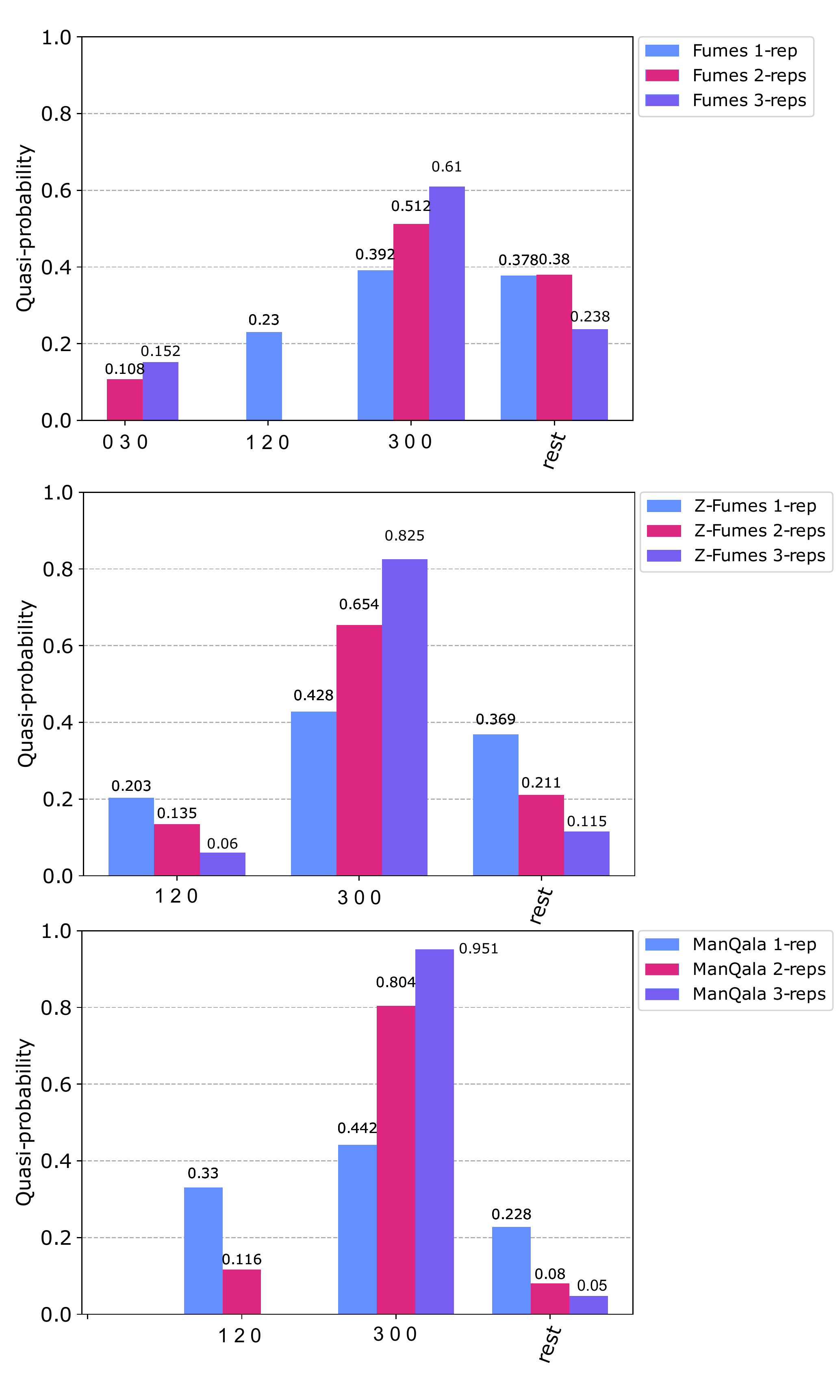}
    \caption{\textbf{Success probabilities of each quantum state engineering strategies under up to 3 repetitions}. The probabilities are calculated in the Schr\"{o}dinger picture after evolving the initial state based on each strategy. Non-deterministic parts of each strategy is repeated for 1, 2, and 3 times using $10,000$ shots each, and averaged over. The labels on the horizontal axis correspond to outcome states of the projective measurements. The label ``rest'' refers to all states that are not the initial and target states.}
    \label{fig:compare3}
\end{figure}

To better understand the impact of repeated applications of each protocol we check the quasi-probability (or the probability density histogram) of the states obtained after the repetition of the protocol multiple times, in order to assess how reliably a protocol steers the quantum state with limited physical \& computational resources. Figure~\ref{fig:compare3} shows the success probabilities of the quantum state engineering strategies under different numbers of repetitions. With a single application, the three strategies have roughly the same probability ($\sim 40\%$) of measuring the target state $\ket{3,0,0}$. However, after 3 repetitions, this probability for ManQala increases to near unity (95\%), in contrast to limited performance increases with FUMES (61\%) and Z-FUMES (83\%). This results from the avoidance of early onset use of measurement-based probabilistic control and dealing with small search spaces when using it in ManQala type strategies. On the other hand, (Z-)FUMES need to re-compute a time duration and apply probabilistic measurement many times. This is tractable for $M=N=3$. However with the growing Hilbert space size, the computational complexity constraints necessitate the usage of sophisticated Tensor Network-based calculations in large computer clusters \cite{orus2019tensor}, and random search is much less reliable.

\section{Generalization of Mancala Inspired Quantum State Engineering}
\label{sec:examples}

\begin{figure}[htb]
    \centering
    \includegraphics[width=0.8\linewidth]{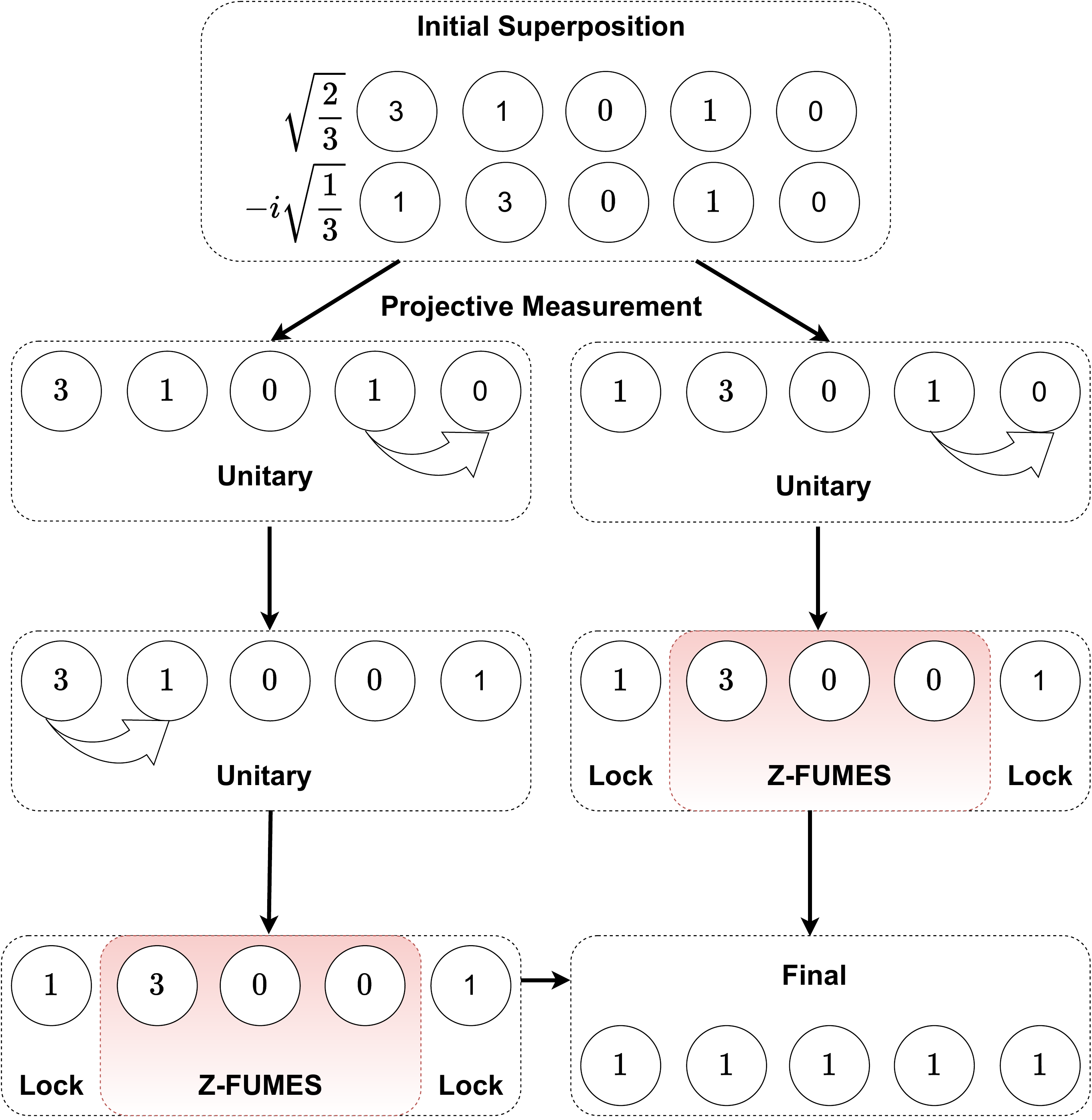}
    \caption{\textbf{An example of the modified ManQala implementation.} A superposition state is projected into one of the two possible states. The target state is demarcated into three sublattices that can be deterministically reachable by one of the possible initial states. Once the sublattices of the possible initial states are driven to the target sublattices, they are Zeno-locked in order to implement Z-FUMES in each sublattice to reach the target state. In the possible trajectories given above }
    \label{fig:m_manqala_ex1}
\end{figure}

In this section, we will give two example implementations of the (modified) ManQala quantum state engineering strategy. Importantly, in this section, we are focusing on the generalized version of ManQala which can begin to significantly diverge from the gameplay of mancala.
In particular, we will show two cases that exemplify how any Fock state, which is not restricted to winnable game configurations, be generated using the algorithm given in Appendix \ref{alg:m_manqala_x}.
In addition, we will loosen our adherence to the mancala game (game rules, direction) in order to further optimize the evolution of the system toward the target state.

The first example has the initial state $\ket{\psi_0}=\left(\sqrt{\tfrac{2}{3}}\ket{3,1}-\tfrac{i}{\sqrt{3}}\ket{1,3}\right)\otimes\ket{0,1,0}$, and the unity occupancy Mott-insulator state as the target state $\ket{\psi_{\text{targ}}}=\ket{1,1,1,1,1}$, on the five site, 1D bosonic lattice (see Figure ~\ref{fig:m_manqala_ex1}). Here the underlying physics is implemented through Bose-Hubbard Hamiltonian with $U=0$ self-interaction potential.

In our case, both the initial and target states have the same average occupancy of $\vb{n}=(1,1,1,1,1)$. However, the former is a superposition state with a high variance in population, while the latter is a product (Fock) state with no variance. Because of this fact, optimizing for the bosonic distance, see the equation (\ref{eq:bosonic_dist}) for $d_B$, in such superposition states is impractical, and mod-ManQala starts by collapsing the initial state right away. We start by doing such a projective measurement to collapse the state into $\ket{3,1,0,1,0}$ with the probability 2/3 or to the state $\ket{1,3,0,1,0}$ with the probability 1/3. In either case, we first designate a target particle number excitation vector of a set of sublattices of the target state. 

In this problem, the nearest reachable, Zeno-lockable set of sublattices of the target state is the one with three sublattices, and with unity particles on the edge and three particles on the middle sublattice (composed of three sites in the middle), $\ket{1,1,1,1,1}\rightarrow \ket{1,{(1,1,1)},1}\rightarrow \ket{1,3,1}$. In other words, the lattice is segmented into $\{1-\text{site},3-\text{site},1-\text{site}\}$ sublattices. For that designated set of sublattices, the former state is $\ket{3,1,0,1,0}\rightarrow\ket{3,(1,0,1),0} \rightarrow \ket{3,2,0}$. While the same sublattice segmentation results in the following sublattice populations for the latter state, $\ket{1,3,0,1,0}\rightarrow \ket{1,(3,0,1),0} \rightarrow \ket{1,4,0}$. 

In order to minimize the bosonic distance between the particle numbers of the target sublattices, $\vb{n}_{\text{targ}}=\left(1,3,1\right)$, and the ones of the states, $\vb{n}_1=\left(3,2,0\right)$ and $\vb{n}_2=\left(1,4,0\right)$ we may choose coherent, deterministic, unitary actions, $\alpha$, that takes the least amount of time. For the former state, one particle needs to hop from the middle sublattice to the right via a two-site unitary evolution of time $\pi/2$ between sites three and four. In other words, an initial projector $\vb{P}_0 =\ket{3}\bra{3}_0\otimes\ket{1}\bra{1}_1\ket{0}\bra{0}_2$ is chosen to acquire a two-site unitary $U_0 = \vb{P}_0\exp\left(-i\pi\vb{P}_0H\vb{P}_0/2\right)$. Then, two particles need to hop from the leftmost sublattice to the middle. This is implemented by another two-site unitary of time $\pi/2$ between sites zero and one, i.e, $\vb{P}_1=\ket{0}\bra{0}_2\otimes\ket{0}\bra{0}_3\ket{1}\bra{1}_4 \rightarrow U_1 = \vb{P}_1\exp\left(-i\pi\vb{P}_1H\vb{P}_1/2\right)$. And, for the latter state, only a single particle needs to hop from the middle sublattice to the right. This is again implemented via a two-site unitary of time $\pi/2$ between sites three and four. Once the bosonic distance $d_B$ between the number vectors of the state sublattices and the target state's sublattice, we can Zeno-lock the sublattices on the edge, then run a Z-FUMES subroutine on the middle sublattice to reach the target state as shown in Figure \ref{fig:m_manqala_ex1}. In the end, a list of (non)unitary actions are taken, $\alpha = \left\{U_0, U_1,\dots, \alpha_{\text{Z-FUMES}}\right\}$. Here $\alpha_{\text{Z-FUMES}}$ at the end of the list corresponds to the set of actions implemented by the latest Z-FUMES step on the $\ket{1,3,0,0,1}$ state, including the two and three site unitaries and projective, global site population measurements.

\begin{figure}[tbh]
    \centering
      \includegraphics[width=\linewidth]{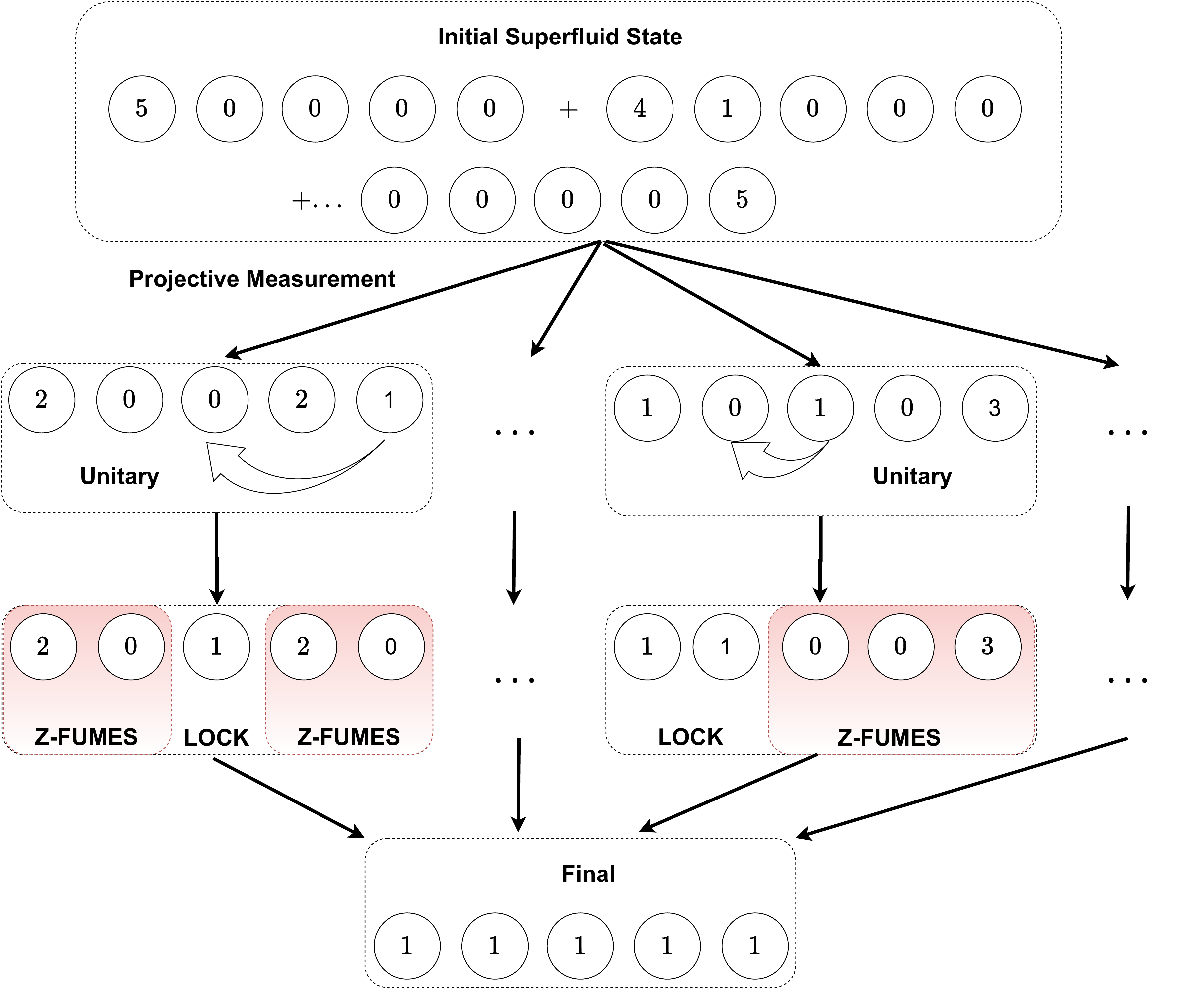}
    \caption{\textbf{An example of the mod-ManQala implementation.} A superfluid state can be projected into many possible states. Here we explicitly show two potential control paths. The target state is demarcated into three sublattices in the left-most example and two sublattices in the right-most example. These sublattice configurations are reachable through deterministic unitary operations that are a function of which state the initial state collapses into upon measurement. Once the sublattices of the possible initial states are driven to the target sublattices, they are Zeno-locked in order to implement Z-FUMES in each sublattice to reach the target state. The example to the left, even though it initially has the correct sublattice populations, starts with a three-site population permutation to place a particle in the very middle. Zeno-locking this site in the middle enables demarcating into three sublattices as the particle in the middle prevents tunneling, which was not possible before as the four sites to the left would act as a four-site sublattice. These ensuing two sublattices to the left and right can be evolved in parallel using Z-FUMES. The example to the right, on the other hand, does not necessitate such caution on placing a demarcating barrier between designated sublattices. Moving the particle in the middle to the left is enough to achieve desired sublattice populations before applying Z-FUMES.}
    \label{fig:m_manqala_ex2}
\end{figure}

Next, we examine a case similar to that analyzed in \cite{sorensen2018quantum} where S{\o}rensen et al. consider a superfluid initial state with $M=N=6$ and steer it towards a Mott-insulator state.
We will consider the $M=N=5$ dimensional version of this same problem, pictured in Figure (\ref{fig:m_manqala_ex2}). 
Note that the initial state is in a superposition while the particle number expectation at all sites is unity.

To perform this control problem using the mod-ManQala algorithm, we again first project the state into one of its constituents either immediately, or after an iteration of FUMES. Then, for each of these possible states, we identify a set of sublattices, to which we minimize the bosonic distance, $d_B$, between the given state and target state's sublattices by moving particles via unitary moves. Once the bosonic hopping distance between the segmented sublattice populations of the target and the pre-processed state is zero $d_B\left(\tilde{P}_{\tilde{\pi}}\vb{n}_0,\vb{n}_{\text{targ}}\right)=0$, we lock the particles in their sites through Zeno-locking (frequent measurements), and run Z-FUMES independently or in parallel in each of these sublattices.
Some possible control paths are pictured in Figure (\ref{fig:m_manqala_ex2}).
One of the possible projected states, $\psi_0=\ket{2,0,0,2,1}$ is split into a two-site sublattice to the left edge, a single site sublattice on the middle, and another two-site sublattice to the right edge, as $\{2-\text{site},1-\text{site},2-\text{site}\}$. As the four-site sublattice on the left already contains the correct number of particles in it, we could have Zeno-locked it and run Z-FUMES immediately. However the search space would be $R=\binom{N+M-1}{N}=\binom{4+4-1}{4}=35$, and it would take much longer for Z-FUMES to converge. On the other hand, when we split it into two sublattices each carrying two particles and two sites, we are searching in parallel within two $R=\binom{2+2-1}{2}=3$ dimensional spaces for the target sublattice using Z-FUMES.

If two sublattices that require further processing by Z-FUMES are side-by-side then there is no barrier to tunneling.  To avoid this we can use the unitary time evolution of the three sites of sites 2, 3, and 4 for a duration of $\sqrt{2}\pi/2$ to move the rightmost particle to the center. This three-site unitary is implemented via the projector, $\vb{P}_0=\ket{2}\bra{2}_0\otimes\ket{0}\bra{0}_1 \rightarrow U_0 = \vb{P}_0\exp\left(-i\sqrt{2}\pi\vb{P}_0H\vb{P}_1/2\right)$. That way the $d_B$ distance between the demarcated sublattice particle numbers of the new state, $\ket{2,0,1,2,0}\rightarrow \ket{2,1,2}$ and the target state $\ket{1,1,1,1,1}\rightarrow \ket{2,1,2}$ is zero, as shown on the left-hand side in Figure (\ref{fig:m_manqala_ex2}). Therefore two Z-FUMES operations can be implemented to the left and rightmost two-site sublattices in parallel. The list of actions for this mod-ManQala scenario would be a concatenation of deterministic permutations (just $U_0$ in this case), and the actions from two Z-FUMES in parallel (one acting left-most, the other right-most sublattices, see Figure \ref{fig:m_manqala_ex2}), $\alpha = \left\{U_0, \alpha_{\text{Z-FUMES-1}}, \alpha_{\text{Z-FUMES-2}}\right\}$.

The last example among these projected Fock states is $\ket{1,0,1,0,3}$. For this example, a $\{1-\text{site},1-\text{site},3-\text{site}\}$ sublattice demarcation is chosen. In that segmentation, the target particle numbers vector is $\vb{n}_{\text{targ}}=\left(1,1,1,1,1\right)\rightarrow\left(1,1,3\right)$ and the state particle number vector is  $\vb{n}_{0}=\left(1,0,1,0,3\right)\rightarrow\left(1,0,4\right)$. To minimize the bosonic distances between the sublattices of these particle numbers $d_b(\vb{n}_{0},\vb{n}_{\text{targ}}; \alpha)$, a two-site unitary of duration $\pi/2$ is chosen between sites 1 and 2 by designating the Zeno-locking projector $\vb{P}=\ket{1}\bra{1}_0\otimes\ket{0}\bra{0}_3\otimes\ket{3}\bra{3}_4$, to move one particle from right most sublattice to the middle. Here the action given after the semi-colon in $d_{b}$, $\alpha = \left\{U = \vb{P}\exp\left(-i\vb{P}H\vb{P}\pi/2\right)\right\}$ represents the site-population permutations applied to the initial state represented by the initial site populations, $\vb{n}_{0}$, such that the distance metric yields the bosonic distance between the transformed and the target states (Figure \ref{fig:m_manqala_ex2}). The remaining actions in this scenario are coming from Z-FUMES acting on the $\ket{0,0,3}$ sublattice, $\alpha = \left\{U, \alpha_{\text{Z-FUMES}}\right\}$.

\end{document}